\documentclass[preprint,showpacs,preprintnumbers,pre]{revtex4}
\usepackage{amssymb}
\usepackage{graphicx}
\usepackage{dcolumn}
\usepackage{bm}

\begin{document}

\title{Alternative approach to community detection in networks}
\author{A.D. Medus}
\email{admedus@gmail.com}
\author{C.O. Dorso}
\email{codorso@df.uba.ar}
\altaffiliation[also member of]{ Carrera del Investigador Cient\'ifico de CONICET}
\affiliation{Departamento de F\'isica, Facultad de Ciencias Exactas y Naturales,
Universidad de Buenos Aires, Pabell\'on 1, Ciudad Universitaria, Ciudad
Aut\'onoma de Buenos Aires (1428), Argentina}
\date{\today }

\begin{abstract}
The problem of community detection is relevant in many disciplines of
science and modularity optimization is the widely accepted method for this
purpose. It has recently been shown that this approach presents a resolution
limit by which it is not possible to detect communities with sizes smaller
than a threshold which depends on the network size. Moreover, it might
happen that the communities resulting from such an approach do not satisfy
the usual qualitative definition of commune, i.e., nodes in a commune are
more connected among themselves than to nodes outside the commune. In this
article we introduce a new method for community detection in complex
networks. We define new merit factors based on the weak and strong community
definitions formulated by Radicchi et al (\textit{Proc. Nat. Acad. Sci. USA} 
\textbf{101}, 2658-2663 (2004)) and we show that this local definitions
avoid the resolution limit problem found in the modularity optimization
approach.
\end{abstract}

\maketitle




\section{\label{sec1}Introduction}

The problem of community detection in complex networks has recently
attracted the attention of researchers in different areas of scientific
knowledge. This is due to the fact that it is a common practice to represent
some complex systems as networks constituted by interconnected nodes.

A network $G$ is defined by a set of nodes \{n\} ($n_{1}$, $n_{2}$,...,$%
n_{n} $.), and a set of links \{l\} ($l_{12},l_{14}...,l_{km}).$ A link $%
l_{ij}$ denotes a relation between node $n_{i}$ and node $n_{j}$. Depending
on the possible values of $l_{ij}$ the resulting network can be of two
types. If $l_{ij}$ can only have the values 1 or 0, we will call the network
unweighted, on the other hand, a network will be defined as weighted if $%
l_{ij}$ can attain values different from 0 or 1, thus indicating that the
relation between nodes is also characterized by a given strength. In this
work we will focus on unweighted networks. We will assume that for every
node $n_{i}$ there exists at least another node $n_{j}$ such that $l_{ij}$
is different from 0, moreover we will consider networks such that for every
conceivable pair of nodes there will be a path (i.e. a sequence of links \{$%
l_{ij}l_{jk}l_{km}...\}$) joining them, in such a case we say that we are
dealing with connected networks. We will consider that the links are
undirected i.e. $l_{ij}=l_{ji}$. Further on, we will focus on sparse
networks for which the number of links in \{l\}, $L$, is much less than the
maximum possible number of links, $L_{\max }$ given by $L_{\max }={N(N-1)}/2$
, with $N$ the total number of nodes in $\{n\}$ .

Generally, complex networks contain a large number of nodes and links, and
it is often possible to decompose them into subgraphs called \textit{%
communities}, selected according to a given criterion. A community is
usually defined, qualitatively, as a subgraph of the network whose nodes are
more connected among them than to nodes outside the subgraph \cite%
{Barthelemy,Newman1}.

Community detection has a wide range of applications. The partition of a
network in communities might allow us to find an specific function naturally
assigned to each community, as for example in the case of metabolic networks 
\cite{Guimera}. On the other hand, community detection can help us to
identify social groups in a social network or can be used to perform a
coarse-graining reduction of the network to simplify subsequent analysis 
\cite{Arenas}.

There are many methods to decompose a network into communities, but the
widely adopted in recent years is the one proposed by Newman and Girvan \cite%
{Newman1}. These authors define a merit factor named \textit{modularity} ($%
Q_{N}$) that quantifies the quality of a given $m$-subgraphs partition $%
\{C_{j}\}_{1\leq j\leq m}$ of the graph $G$, where $\bigcup_{j=1}^{m}C_{j}=G$
with $C_{i}\bigcap C_{j}=\varnothing $ if $i\neq j$. This quantity measures
the difference between the actual fraction of internal links in each
subgraph with respect to the expected value of the same quantity, if nodes
in the network are randomly connected keeping the degree of each one fixed.
The best partition of the network is taken as the one that maximizes the
modularity $Q_{N}$, in this way, the network partition problem is turned
into an optimization one.

Modularity optimization is a hard problem due to the fact that the number of
possible partitions of a network increases at least exponentially with its
size. Indeed, it has recently been proven that this problem is NP-complete 
\cite{Brandes}, and then, there is not a correct polynomial-time algorithm
to solve it for networks of any size. Many optimization algorithms have been
developed, like simulated annealing \cite{Guimera,MAD}, extrema optimization 
\cite{Duch} and spectral division \cite{Newman2}, but all of them can only
give an approximation to the optimum partition for large networks.

In this work, we do not introduce a new $Q_{N}$ optimization algorithm, but
we propose new merit factors for the calculation of the partition of
networks into communities.

The modularity $Q_{N}$ is a \textit{non-local} community definition in the
sense that it is necessary to know general characteristics of the whole
network in order to decide if a given subgraph of the network is a
community. In a recent paper \cite{Barthelemy} Fortunato $\&$ Barth\'{e}lemy
have shown that this non-local character imposes a resolution limit, by
which the minimal community size that can be detected, by modularity
optimization, depends on global network parameters. Then, $Q_{N}$
optimization is not able to detect communities of size smaller than a given
threshold.

In this work we use the weak and strong community definitions proposed by
Radicchi et al in \cite{Radicchi}. We emphasize their local character and
introduce new merit factors to evaluate the quality of a given partition of
a network based on these quantitative definitions of community. Then, we
implement an optimization method in the spirit of simulated annealing \cite%
{MAD,ecra}, in order to analyze different networks using these new merit
factors and to show the characteristics of our approach. Finally, we show
that the resolution limit problem does not appear in our approach.

The paper is organized as follows. In section \ref{sec2} we review the
definitions introduced in \cite{Radicchi} and compare them with the
modularity $Q_{N}$, we also analyze the meaning of this last quantity. In
section \ref{merit} we define the \textit{community strength} $S$ in the
strong and weak sense and we introduce the associated merit factors. In
section \ref{sec3} we apply our method to different, well known, networks.
In section \ref{sec4} we analyze the resolution limit problem for our
approach. Finally, conclusions are drawn in section \ref{sec5}.

\section{\label{sec2}Community definitions}

When thinking about communities in networks we have in mind a qualitative
community definition:\textit{\ a community is a group of nodes in which the
number of internal links, connecting nodes within the group, is larger than
the number of external ones}. In order to formalize this qualitative
criterion we consider a graph $G$ containing $N$ nodes, with $k_{i}$ the
degree of node $i\in G$. If $C$ is a subgraph of $G$ with $k_{i}^{in}$ and $%
k_{i}^{out}$ the number of links of node $i\in C$ that connect it to nodes
inside and outside of $C$ respectively. There are two quantitative community
definitions introduced by Radicchi et al \cite{Radicchi}: \newline

\textbf{i) Community in strong sense:} \textit{$C$ is a community in the
strong sense if:} 
\begin{equation}
k_i^{in}>k_i^{out}\:\:\: \forall i\in C
\end{equation}

\textbf{ii) Community in weak sense:} \textit{$C$ is a community in weak
sense if:} 
\begin{equation}
\sum_{i\in C}k_i^{in}>\sum_{i\in C}k_i^{out}
\end{equation}

In words: a subgraph $C\subset G$ will be a community in the strong sense if
each of its nodes has more links connecting it with nodes in $C$ than those
that connect it with other nodes not belonging to $C$. In a similar way, $%
C\subset G$ will be a community in the weak sense if the sum of the number
of links that interconnect nodes inside $C$ is larger than the sum of all
links that connect nodes in $C$ with nodes not belonging to $C$. These
community definitions are simple, intuitive and \textit{local}: given a
subgraph $C\subset G$ we can decide if it constitutes a community, in either
strong or weak sense, without knowledge of the entire structure of $G$.

In order to compare the previous approach with the one proposed in \cite%
{Newman1}, we briefly review the definition and meaning of the modularity $%
Q_{N}$. Given a $m$-subgraphs partition $\{C_{j}\}_{1\leq j\leq m}$ of the
graph $G$, where $\bigcup_{j=1}^{m}C_{j}=G$, the mathematical expression of $%
Q_{N}$ is : 
\begin{equation}
Q_{N}=\sum_{i=1}^{m}\left[ \frac{l_{i}}{L}-\left( \frac{d_{i}}{2L}\right)
^{2}\right]  \label{eq1}
\end{equation}%
where $l_{i}$ denotes the total number of internal links for subgraph $%
C_{i}\subset G$ and $d_{i}=\sum_{j\in C_{i}}k_{j}$, and $L=\frac{1}{2}%
\sum_{j\in G}k_{j}$ is the total number of links in $G$.

The term $l_{i}/L$ in Eq. \ref{eq1} denotes the actual fraction of internal
links in subgraph $C_{i}$, while $d_{i}/2L$ can be interpreted as the
probability of a link to be connected to a node in subgraph $C_{i}$. Then, $%
\left( d_{i}/2L\right) ^{2}$ is the expected fraction of links within
subgraph $C_{i}$ when all nodes in $G$ are randomly connected keeping the
degree of the nodes fixed. This last ideal random picture is used to compare
with the actual one because it is assumed that corresponds to a situation
with no communities (although it was shown in \cite{Guimera2} that random
networks may have a community structure).

As already mentioned, the modularity $Q_{N}$ was conceived as a measure of
the goodness of a given partition of the network. Then, the bigger $Q_{N}$
is, the better the partition is. We should notice that this merit factor
implies, in turn, a community definition (which does not necessarily
corresponds to the intuitive one stated above): a subgraph $C_{j}$ will be a
community if the actual number of links that connects nodes in $C_{j}$ is
bigger than the expected one when all nodes in the network are randomly
connected, this is to say, when $l_{i}/{L}-\left( d_{i}/{2L}\right) ^{2}>0$.
Clearly this last condition depends on the global parameter $L$ , then, we
say that the community definition associated with $Q_{N}$ is \textit{%
non-local}.

In what follows, we will introduce new merit factors associated with the
weak and strong community definitions.

\subsection{\label{merit}Merit factors for weak and strong\newline
community definitions. Community strength}

Given a graph $G$ and a $m$-subgraphs partition $\{C_{j}\}_{1\leq j\leq m}$,
where each subgraph $C_{j}\subset G$ constitutes a community according to
any of the local definitions mentioned in the previous section, we want to
define a quantity that measures the \textquotedblleft
quality\textquotedblright\ of each of the resulting communities. In the
context of the above mentioned local framework, this quantity must only
depend on the local characteristics of the subgraph $C_{j}$. Therefore, our
analysis must be circumscribed to nodes and links belonging to $C_{j}$ and
external links that connect nodes in $C_{j}$ to nodes outside $C_{j}$.
Following the weak and strong definitions of community, the more internal
links a community has, with respect to the external ones, the
\textquotedblleft stronger\textquotedblright\ it will be. If $%
k_{i}=k_{i}^{in}+k_{i}^{out}$ is the degree of node $i\in C_{j}$, where $%
k_{i}^{in}$ and $k_{i}^{out}$ are the number of internal and external links
for node $i$, we define the \textit{\textquotedblleft community
strength\textquotedblright } ($S$), that measures the normalized difference
between internal and external links for nodes in $C_{j}$:

\begin{equation}
S(C_{j})=\sum_{i\in C_{j}}\frac{k_{i}^{in}-k_{i}^{out}}{2L(C_{j})}
\label{eq2}
\end{equation}%
were $L(C_{j})=\frac{1}{2}\sum_{i\in C_{j}}k_{i}$. Then, $-1\leq
S(C_{j})\leq 1$, and it achieves its maximum value $1$ when $k_{i}^{out}=0$
\ $\forall i\in C_{j}$.

The definition of $S(C_{j})$ according to Eq. \ref{eq2} is valid for
unweighted networks. In the case of weighted links, we have to interpret $%
k_{i}$ as the sum of the weights of the links that connect to node $i$, for
both $k_{i}^{in}$ and $k_{i}^{out}$.

Now we introduce a merit factor $Q_{W}$ for the weak community definition as
the sum of $S(C_{j})$ over all subgraphs $C_{j}\subset G$:

\begin{equation}
Q_{W}=\sum_{j=1}^{m}S(C_{j})=\sum_{j=1}^{m}\sum_{i\in C_{j}}\frac{%
k_{i}^{in}-k_{i}^{out}}{2L(C_{j})}  \label{eq3}
\end{equation}

with the constraint that each subgraph $C_{j}\subset \{C_{j}\}_{1\leq j\leq
m}$ must satisfy the weak community definition i.e.

\begin{equation}
S(C_{j})>0\text{ }\forall C_{j}\subset \{C_{j}\}_{1\leq j\leq m}
\end{equation}

As in the case of $Q_{N}$: the bigger $Q_{W}$ is, the better the $m$%
-subgraphs partition $\{C_{j}\}_{1\leq j\leq m}$ of $G$ will be, in the
sense of weak community definition. Then, it is possible to implement the
optimization algorithms developed for $Q_{N}$ for this new merit factor $%
Q_{W}$,

In Eq. \ref{eq3}, $Q_{W}=1$\ when all the network constitutes a single
community. If as a result of the optimization process the maximum value
obtained for $Q_{W}$ is precisely $Q_{W}=1$ and we get a single community,
then the best partition of the graph corresponds to no partition. However,
it is possible that one could get $S(C_{j})>0$\ $\forall C_{j}\subset
\{C_{j}\}_{1\leq j\leq m}$\ for a given $m$-subgraphs partition, with $m>1$
but with $0<Q_{W}<1$. The resulting community structure of network would
correspond to a suboptimal partition.\bigskip

In the same spirit we now define a merit factor $Q_{S}$ according to the
strong community definition:

\begin{equation}
Q_{S}=\sum_{j=1}^{m}S(C_{j})=\sum_{j=1}^{m}\sum_{i\in C_{j}}\frac{%
k_{i}^{in}-k_{i}^{out}}{2L(C_{j})}
\end{equation}%
with the constraint 
\begin{equation}
(k_{i}^{in}-k_{i}^{out})>0 \:\forall i\in C_{j}
\end{equation}

Now, our definition of optimal partition can be stated in the following way:

\textbf{Definition:} \textit{the optimal $m$-subgraphs partition $\{C_{j}\}_{1\leq j\leq m}$
 of a graph $G$ in the Strong (Weak) sense is that one with maximal merit factor} $Q_{S}$ \textit{(}$Q_{W}$\textit{)}.

In next section we will show some examples of the application of this new
merit factors in network partition problems.

\section{\label{sec3}Examples}

In all examples presented in this section we have used an optimization
algorithm based on simulated annealing, described in \cite{MAD}, but for our
new merit factors. The optimization can be performed in two ways. In the
first one the total number of communes is left as a free parameter and as a
consequence the final number of communes is determined by the simulated
annealing process. In the second one, the number of communes is taken as an
extra constraint. We will always use the first approach unless it is
explicitly stated that the number of communes is fixed. This last
methodology might be used when the optimal number of communes is already
known from the experiments as in the case of the Zachary network.

\subsection{Zachary's karate club network.}

\begin{figure}[tbp]
\includegraphics[scale=1.5]{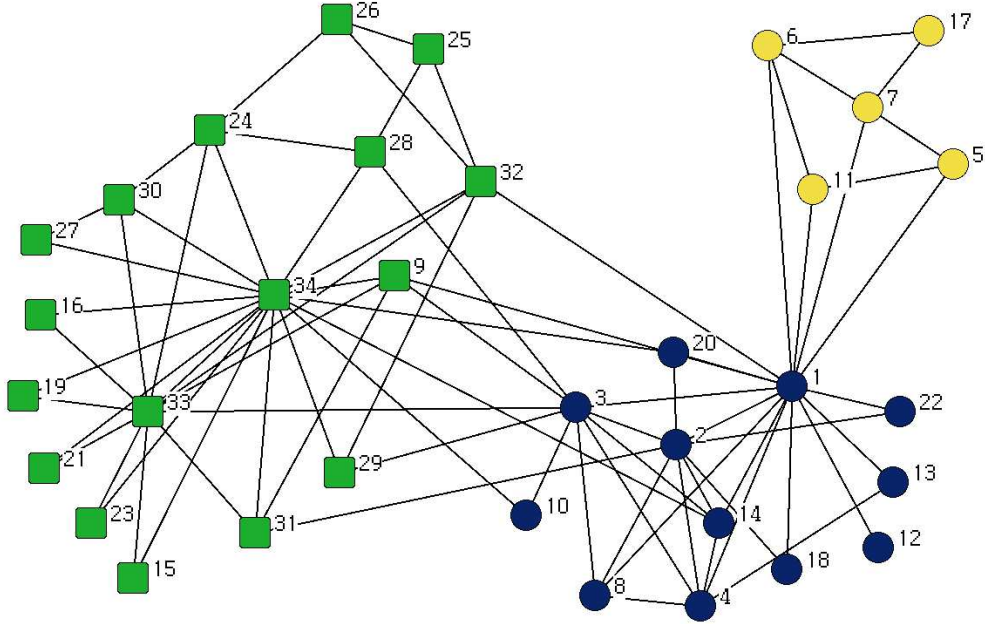}
\caption{(Color online) Best partition for Zachary network. Squares and
circles denote the two communities obtained with our approach when the
number of communities is fixed to two. This partition correspond
to the one consigned by Zachary in \protect\cite{Zachary}, with exception of node $10$ that appear misclassified.}
\label{fig1}
\end{figure}

We will begin with a typical case: Zachary's Karate Club \cite{Zachary},
that has turned into an unavoidable example in publications about community
structure. This network represents the relationships between members of a
karate club at a University in the 1970s, and it has been shown that it has
a strong community structure in many previous studies \cite{Newman1,MAD}.
Applying the optimization algorithm for the weak community definition merit
factor $Q_{W}$, we obtained, for the unweighted version of Zachary network,
a partition into three communities of sizes: $17$ ($C_{1}$), $12$ ($C_{2}$)
and $5$ ($C_{3}$) nodes, with a value of $Q_{W}=1.792$ (Fig. \ref{fig1}).
When the number of communities was constrained to two, we obtained two communities of 17 nodes
each one, with $Q_{W}=1.487$ (circles and squares in Fig. \ref{fig1}). This partition corresponds to the one observed by Zachary with the exception of node $10$ that appear misclassified.

With this analysis we can know, in addition, the strength $S(C_{j})$ of each
community $C_{j}$ in the network. For the best partition of the Zachary
network into three communities of $17$ ($C_{1}$), $12$ ($C_{2}$) and $5$ ($%
C_{3}$) nodes, we have: $S(C_{1})=0.744$ , $S(C_{2})=0.548$ and $%
S(C_{3})=0.5,$ with $C_{1}$ as the strongest community. On the other hand,
the partition into two communities, is composed by two strong communities of 
$17$ nodes each one, with $S(C_{j})=0.744$ for each one.

When we perform the community analysis using the strong community merit
factor $Q_{S}$, we obtained two communities: $C_{1}$ with $29$ nodes ($%
S(C_{1})=0.943$) and $C_{2}$ with $5$ nodes ($S(C_{2})=0.5$). In Fig. \ref%
{fig1} can be observed that node 10 has one internal and one external link
and this situation is not be allowed in the strong community definition. For
this reason, the communities with $17$ and $12$ nodes are joined together.

\subsection{Star network.}

\begin{figure}[tbp]
\includegraphics[scale=1.5]{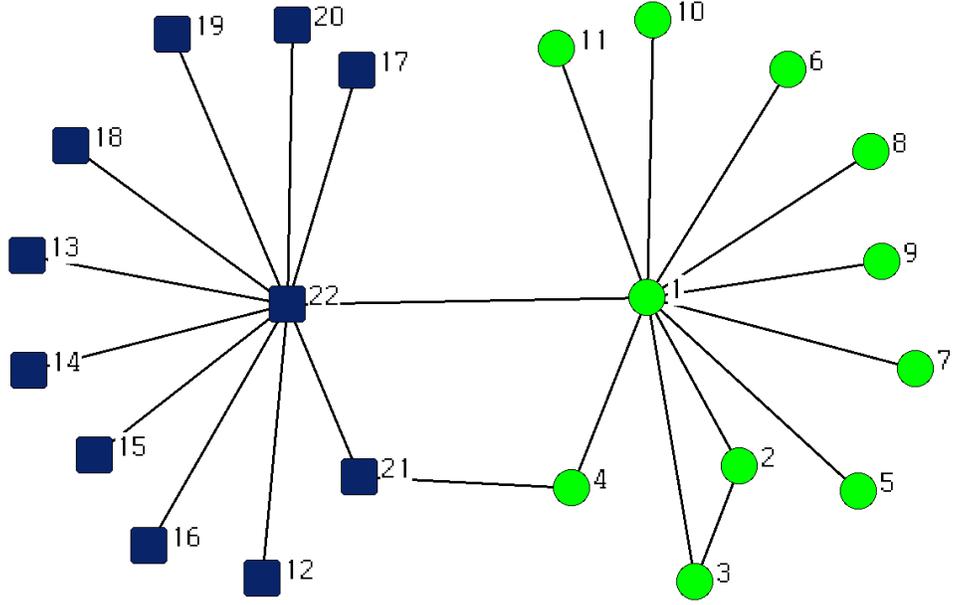}
\caption{(Color online) Community structure for star network obtained in our approach. The
two communities are distinguished with squares and circles. For $Q_N$
optimization we find an additional community containing nodes $4$ and $21$
with strength $S=0$.}
\label{fig2}
\end{figure}

Another testing example is the star network of Fig. \ref{fig2}, consisting
of two interconnected stars with $11$ nodes each one. In the weak community
picture we obtain for the optimal community structure a partition into two
communities of $11$ nodes each with a value of $Q_{W}=1.652$. The
corresponding strengths are $S(C_{1})=0.833$ (circles in Fig. \ref{fig2})
and $S(C_{2})=0.818$ (squares in Fig. \ref{fig2}). The difference in
strength between the two communities is ascribed to the extra link that
connects nodes $2$ and $3$.

We must notice that, in the same context, there is another partition with
the same value of $Q_{W}$ but composed by two communities with $12$ and $%
10$ nodes. This happens when node $4$ is moved from one community to the
other in Fig. \ref{fig2}.

No partition was obtained when we use the strong community merit factor $%
Q_{S}$. This is due the fact that nodes $4$ and $21$ are singly connected
and then the condition $k_{i}^{in}-k_{i}^{out}>0$ is not satisfied.

When $Q_{N}$ optimization was implemented, we obtained three communities: $%
C_{1}$ and $C_{2}$ with $10$ nodes each and $C_{3}$ with $2$ nodes,
including nodes $4$ and $21$, with strength $S(C_{3})=0$ which does not
satisfy any of the quantitative community definitions reviewed in \ref{sec2}.

\subsection{Ring network.}

\begin{figure}[tbp]
\includegraphics[scale=1.5]{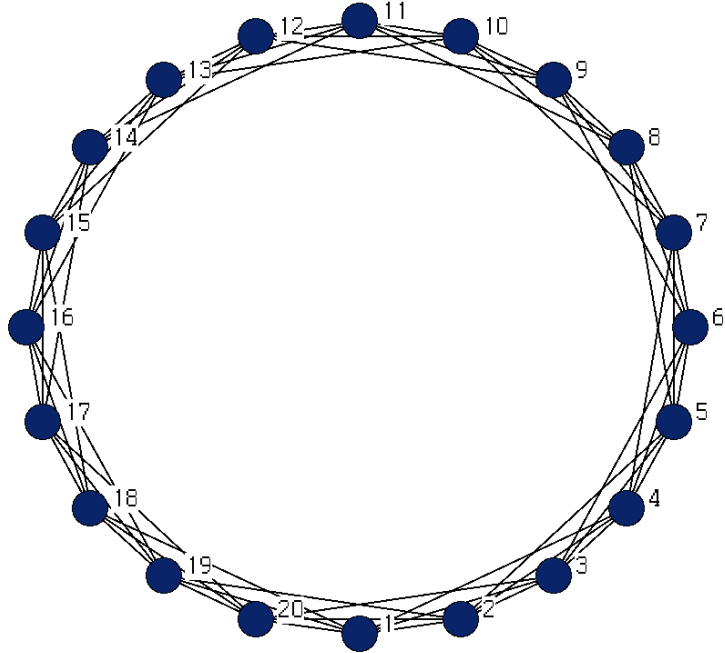}
\caption{(Color online) Ring network. This symmetric network does not present an intrinsic
community structure. However, almost all approaches for community detection
find a community structure for this example, but the identities of the nodes
within each community change when the detection process is repeated.}
\label{fig3}
\end{figure}

Another example is the ring network of Fig. \ref{fig3} with $20$ nodes and $%
k=6$. This network can not have a community structure due to its symmetry.
However, we have obtained two communities of $10$ nodes each one, with
strength $S=0.6$ by means of the weak merit factor optimization. We must
notice here, that the found communities are not unique, that is to say, on
having applied repeatedly the algorithm different communities of the same
size but involving different sequences of node indexes are obtained. This is
an evidence of the absence of an underlying community structure. This
unsatisfactory result is also obtained when optimizing the $Q_{N}$ merit
factor, but in this case the optimal partition is into $3$ communities, two
of them with $7$ nodes each, and the third one with $6$ nodes

On the other hand, when we run the optimization algorithm with the strong
community condition, no partition is obtained. This is true for all ring
networks because an hypothetical frontier node, with $k_{i}^{in}=k_{i}^{out}$%
, will not satisfy the strong community condition.

\subsection{The bottlenose dolphins network}

\begin{figure}[tbp]
\includegraphics[scale=1.5]{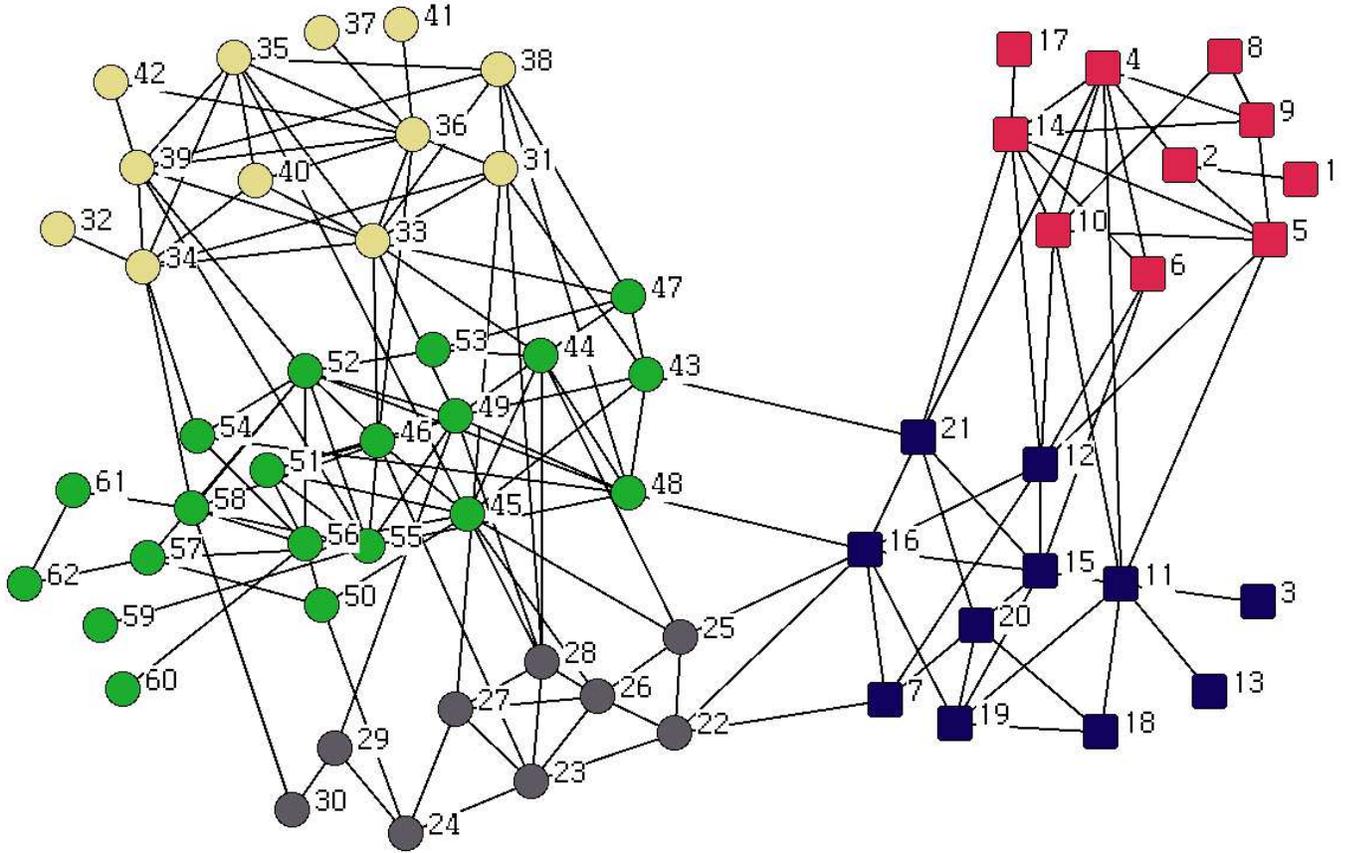}
\caption{(Color online) Bottlenose dolphin network. This network has a size of 62 nodes and
it is known from direct observation that it has two communities. In this
figure squares and circles denote the communities detected by our Strong
community approach and the colors (shades of gray or colors on line) show
the results of the Weak community approach. Notice that the optimization
according to $Q_{W}$ merely subdivides the communities obtained through $Q_{S}$ optimization.}
\label{fig4}
\end{figure}

Another social network which has attracted considerable interest is the one
corresponding to the bottlenose dolphins network, which has been fully
analyzed in \cite{dolphin} (see also \cite{Newman1}). This small social
network is composed by 62 nodes and it is known to consist of two
communities of sizes 41 and 21 nodes each. Following the approach proposed
in this work we first analyzes this network applying the Q$_{N}$ analysis in
our simulated annealing approach. The result of this analysis is the
partition of the network into four communities composed by $21$, $16$, $13$,
and $12$ nodes each. When we performed the optimization of the Weak
community definition we obtained five communities of $20$, $12$, $11$, $10$
and $9$ nodes each. Finally when the dolphin network is analyzed in terms of
the strong community definition we obtained the actual partition, as
observed experimentally, in two communities of $41$ and $21$ nodes each.
These last two results are displayed in Fig. \ref{fig4}. In this figure we
show the two communities according to the Strong community definition as
circles ($41$ nodes community) and as squares ($21$ nodes community). The
corresponding analysis according to the Weak community definition further
divides the previous two communities and are denoted by the different shades
of gray (see caption for details) in the figure. It should be noted at this
point that when the optimization of $Q_{W}$ is performed with the extra
constraint that the number of communes is $2$ we obtain the same community
structure as observed experimentally.

\subsection{Computational generated test network.}

We conclude our short list of examples with the analysis of computer
generated graphs which have a community structure. A word of caution should
be raised at this point, because the communities built into these graphs are
usually of uncertain nature and being purely theoretical, the assumed
community structure cannot be \textquotedblleft verified experimentally\textquotedblright as in the case of
the Zachary Karate Club or the bottlenose dolphins case.

\subsubsection{Methodology I}

In this case we use the method proposed and analyzed in \cite{Duch,Newman3}.
We take a $128$ nodes graph $G$ divided in four modules $C_{j=1,...,4}$ of $%
32$ nodes each and with nodes degree $k_{i}=k_{i}^{in}+k_{i}^{out}=16$ $%
\forall i\in G$. We have earlier defined $k_{i}^{in}$ ($k_{i}^{out}$) as the
number of links that connect node $i\in C_{j}$ to another node in (out of) $%
C_{j}$. When $k_{i}^{out}$ is varied between $0$ to $16$, $G$ go from a
strong communality graph to a quasi-random one.

In the framework of weak and strong merit factors optimization the expected
four communities partition was obtained for $0\leq k_{i}^{out}\leq 7$, where 
$Q_{W}<1$ only for $k_{i}^{out}=7$. When $k_{i}^{out}\geq 8$ the mean number
of external links is bigger or equal to internal ones in each module and no
partition was obtained.

\subsubsection{Methodology II}

We now use the formalism introduced in \cite{Fortunato}, in which an
algorithm for generating a class of benchmark graphs, that account for the
heterogeneity in the distributions of node degrees and of community sizes
was devised. The aim of this algorithm is to built a graph with more or less
well defined community structure (in the caption of Figure $5$ in this paper a
reference is made to \textquotedblleft communities in the strong sense\textquotedblright which is not the case
for this algorithm as is easily verified). It is assumed that both the degree
and the community size distributions are power laws, with exponents $\gamma$ and $\beta$, respectively. The number of nodes is $N$, the average degree is $%
<k>$. One more parameter characterizing this model is the mixing parameter $%
\mu $. Each node shares a fraction $1-\mu$ of its links with the other
nodes of its community and a fraction $\mu$ with the other nodes of the
network.

We have generated graphs according to this algorithm and we have analyzed
them using the Girvan-Newman definition of community using our global
optimization approach \cite{MAD} and the one proposed in this
work. The parameters defining the graphs were chosen to be: $\gamma=2.5,$ $%
\beta=1.5,$ $0.1\leq \mu \leq 0.6$. The size of the graphs was fixed in $%
300$ nodes, and the mean degree in $8$ (maximum degree=$30$).

\begin{figure}[tbp]
\includegraphics[scale=0.5]{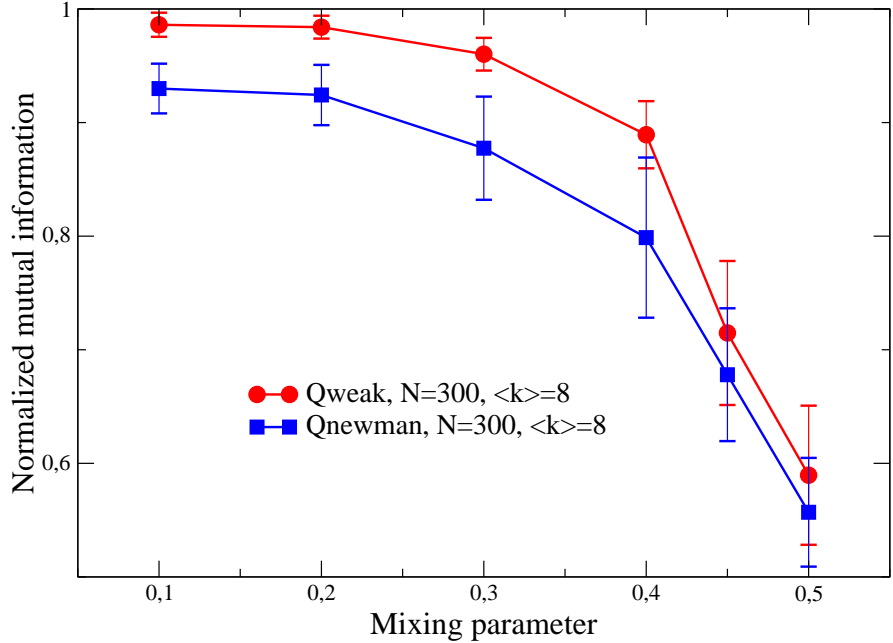}
\caption{(Color on line) Computer generated test network. We have generated graphs according to the formalism proposed by Lancichinetti et al. The
results of our calculation are displayed by the full circles (in red) while
the results according to the Girvan-Newman approach are denoted by square (in blue).}
\label{fig10}
\end{figure}

The results of such a calculation are summarized in Fig. \ref{fig10} (see
captions for details). The quality of the partitions obtained with the
recognition algorithms with respect to the communities established by the
algorithm of Lancichinetti et al. is measured by means of the so called
\textit{Normalized Mutual Information} \cite{NMI}. According to the results
displayed in Fig. \ref{fig10}, our merit factors definitions outperforms the Girvan-Newman approach.

\section{\label{sec4}Resolution limit problem}

In a recent paper \cite{Barthelemy} S. Fortunato and M. Barth\'{e}lemy
showed that modularity $Q_{N}$ optimization fails to detect communities
smaller than a certain threshold which depends on global parameters of the
network under study, as is the case of the total number of links in the
network.

Following \cite{Barthelemy} we can define a community from Eq. \ref{eq1}, in
the framework of modularity $Q_N$, as a subgraph $C_i\subset G$ that satisfy:

\begin{equation}  \label{eq4}
\frac{l_i}{L}-\left(\frac{d_i}{2L}\right)^2>0
\end{equation}

This expression can be interpreted as the community strength in this
framework. We can write $d_{i}=2l_{i}+l_{i}^{out}$, where $l_{i}$ and $%
l_{i}^{out}$ denote the number of internal and external links for subgraph $%
C_{i}$, and write $l_{i}^{out}$ as a fraction of internal links $%
l_{i}^{out}=a l_{i}$ with $a\geq 0$. Then, from Eq. \ref{eq4}, the following
condition for the community size $l_{i}$ is obtained: 
\begin{equation}
l_{i}<\frac{4L}{(a+2)^{2}}  \label{eq5}
\end{equation}%
The dependency on total number of links $L$ in Eq. \ref{eq4} clearly shows
that the community definition, in the context of modularity $Q_{N}$, is
non-local. In \cite{Barthelemy} the authors show that this non-locality is
the origin of \textit{limit resolution problem}. 
\begin{figure}[tbp]
\includegraphics{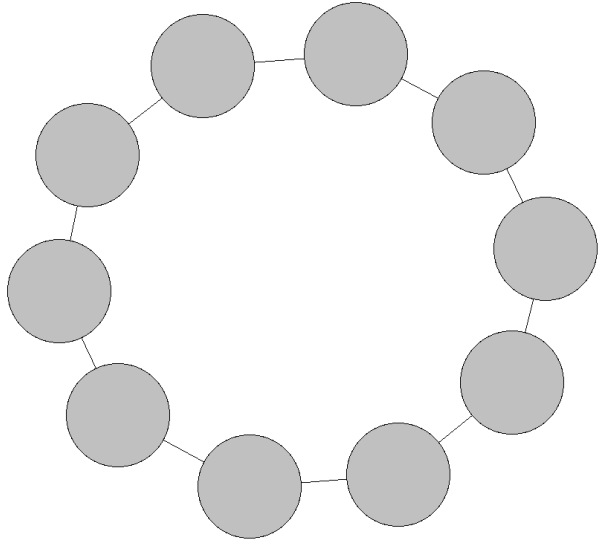}
\caption{Ring of cliques. Each circle represents a clique which is a totally
connected subgraph with $n$ nodes.}
\label{fig9}
\end{figure}

On the other hand, in our approach the community definition is strictly
local. Then, we can decide if a subgraph is a community without regards to
the size of the entire network. To illustrate this conclusion we work out an
example introduced in \cite{Barthelemy}.

Let us suppose a ring of totally connected subgraphs (from now on \textit{%
cliques}) in Fig. \ref{fig9}. Each subgraph has $n$ nodes connected by $%
n(n-1)/2$ internal links and two external ones, and we have $M$ of this
subgraphs with a total number of links $L=M(n(n-1)/2+1)$. The optimal
partition for strong and weak community definition frameworks is the natural
one: each clique constitutes a single community. This can be easily shown by
analyzing another alternative partition in which each community $C_{j}$
contains $W\geq 2$ cliques.

First, we calculate the strength $S(C_{j})$ for one of this subgraphs
containing $W$ cliques: 
\begin{equation}
S(C_{j})=\frac{W(n-2)(n-1)+2(W-1)n+2(n-2)}{W(n-1)n+2W}  \label{eq7}
\end{equation}%
Now, we want to compare the result of Eq. \ref{eq7} with the total strength
of the same subgraph $C_{j}$ when each clique is taken as a single
community. In this case, the strength of $C_{j}$ is given by $W$ times the
strength of one clique: 
\begin{equation}
S^{\ast }(C_{j})=W\frac{(n-2)(n-1)+2(n-2)}{n(n-1)+2}  \label{eq8}
\end{equation}

Then, it is straightforward to see that $S^{\ast }(C_{j})>S(C_{j})$ is
equivalent to: 
\begin{equation}
W(n-2)[(W-1)(n-1)+2W]-2(nW-2)>0  \label{eq9}
\end{equation}

We have said that each clique constitutes a single community in the strong
(and then, also in the weak) community definition, therefore $n\geq 3$.
Then, the condition of Eq. \ref{eq9} is satisfied for all $W\geq 2$. This is
to say that the optimal partition in our approach is that one for which each
clique constitutes a single community. This is not the general case in the
modularity $Q_{N}$ framework. It was showed in \cite{Barthelemy} that, due
to resolution limit problem, partitions in communities with two or more
cliques can give larger values of $Q_{N}$ than with single clique
communities.

\section{\label{sec5}Conclusions}

In this work we have proposed new merit factors to recognize communities in
networks. These merit factors are more realistic than the ones currently in
use in the literature because they strictly adhere to what a community is
expected to be, i.e., a subset of nodes which are more connected among
themselves than to the rest of the network under consideration.

We started by putting forward this qualitative definition of a community and
then we reviewed the meaning of the quite popular measure of the quality of
a given partition known as the modularity $Q_{N}$. As we have discussed
above, the community definition associated to this quantity is non-local and
does not necessarily corresponds to the aforementioned qualitative
definition. One of the consequences of the non-local character intrinsic to
this quantity is the limit resolution problem as stated in \cite{Barthelemy}.

In order to recognize communities in networks that strictly adhere to the
qualitative definition, we have used (following \cite{Radicchi}) two local
community definitions: the weak one and the strong one . In order to use
this definitions to recognize communities we have developed a criteria to
quantify the strength of a community ($S$). Afterwards, we have defined two
merit factors associated with $S$ which we named $Q_{S}$ and $Q_{W}$. As
with $Q_{N}$ the problem of recognizing communities in a network is mapped
onto an optimization problem, i.e., the communities in a network are the
elements of the partition which maximizes $Q_{S}$ or $Q_{W}$. We have
performed the optimization of these merit factors on some standard networks
by implementing an algorithm in the spirit of simulated annealing. The limit
resolution intrinsic to the $Q_{N}$ definition is not present in our
approach.

It is worth noticing at this point that the solution to the detection of
communities in the strong sense is also a solution in the weak sense but not
necessarily optimal. On the other hand, the converse is generally not true
as we stated in section \ref{sec3}.

The strong community definition tends to give larger communities because of
its inability to deal with nodes that are equally shared by two highly
connected subgraphs, but on the other hand has the nice property that it is
the only one that gives no partition for symmetric string networks and also
solves the problem of the bottle nose dolphins network exactly, without
constraints in the number of communities.

\begin{acknowledgments}
C.O.D acknowledge partial support from CONICET through grant PIP5969.
\end{acknowledgments}

\end{document}